\documentclass[12pt]{article}

\usepackage[T1]{fontenc}
\usepackage[utf8]{inputenc}
\usepackage{amsmath,amsfonts,amsthm,amssymb}
\usepackage{booktabs,tabularx,subcaption}
\usepackage{float}
\usepackage{xcolor}
\usepackage{enumerate}
\usepackage{graphicx,color}
\usepackage{hyperref}

\graphicspath{{./figures/}} 

\newcommand{\ignore}[1]{}

\usepackage{booktabs}
\usepackage{siunitx}

\title{Hepatitis C Virus Genotyping with a Transformer Neural Network}

\author{Ariella Aro\renewcommand{\thefootnote}{\arabic{footnote}}\footnotemark[1]
\textsuperscript{,}\renewcommand{\thefootnote}{\fnsymbol{footnote}}\footnotemark[1] \and 
Taim\'a Furuyama\renewcommand{\thefootnote}{\arabic{footnote}}\footnotemark[2] \and 
Marcelo R. S. Briones\renewcommand{\thefootnote}{\arabic{footnote}}\footnotemark[3] \and 
Luis M\'ario R. Janini\renewcommand{\thefootnote}{\arabic{footnote}}\footnotemark[4]
\and 
Isabel M. V. Guedes de Carvalho\renewcommand{\thefootnote}{\arabic{footnote}}\footnotemark[5]
\and 
Fernando Antoneli\renewcommand{\thefootnote}{\arabic{footnote}}\footnotemark[3]
\textsuperscript{,}\renewcommand{\thefootnote}{\fnsymbol{footnote}}\footnotemark[1]
}

\date{\today}

\begin{document}

\maketitle

\renewcommand{\thefootnote}{\arabic{footnote}}
\footnotetext[1]{Departamento de Ciência da Computação, Instituto de Matemática, Estatística e Ciência da Computação, Universidade de São Paulo, São Paulo, SP, Brazil}
\footnotetext[2]{Independent Researcher, S\~ao Paulo, SP, Brazil}
\footnotetext[3]{Centro de Bioinform\'atica M\'edica, Universidade Federal de S\~ao Paulo, S\~ao Paulo, SP, Brazil}
\footnotetext[4]{Departamento de Microbiologia, Imunologia e Parasitologia, Departamento de Medicina, Laborat\'orio de Retrovirologia, Universidade Federal de S\~ao Paulo, S\~ao Paulo, SP, Brazil}
\footnotetext[5]{Laborat\'orio de Virologia, Instituto Butantan, S\~ao Paulo, SP, Brazil}

\renewcommand{\thefootnote}{\fnsymbol{footnote}}
\footnotetext[1]{Correspondence: \href{mailto:ariella@ime.usp.br}{ariella@ime.usp.br}, \href{mailto:fernando.antoneli@unifesp.br}{fernando.antoneli@unifesp.br}}

\begin{abstract} 
This study aims to explore the applicability of Transformer-based models for genetic sequence classification by evaluating their performance in predicting hepatitis C virus (HCV) genotypes and subtypes after fine-tuning. A total of 2,881 HCV whole-genome sequences obtained from the Los Alamos HCV Sequence Database were used, including genotypes 1 to 6 and all confirmed subtypes. Genotypes 7 and 8 were excluded due to an insufficient number of samples.
The fine-tuning process was based on several datasets that differed in fragmentation method, data volume per file, and labeling. In genotype classification, fine-tuning strategies employing homogeneous fragmentation and balanced sample distribution resulted in higher performance, with precision ranging from 98.48\% to 100\%. In contrast, fine-tuning conducted using a fragmentation strategy that caused data imbalance, along with an arbitrary distribution of samples across training files, achieved a precision of 48.12\%, which is considered low compared with other models.
This configuration, which was also manually evaluated, resulted in a high error rate in genotype 5 prediction due to its low frequency in the datasets used. In subtype classification, the best-performing fine-tuning approach achieved 99.89\% accuracy and 99.87\% precision. Models that included additional genotypes showed a slight decrease in performance due to the increased complexity of the task.
This study demonstrates that, when fine-tuning datasets contain properly fragmented, distributed, and labeled genetic sequences, Transformer-based neural networks can achieve high performance and are a promising approach for HCV genotype and subtype classification.

\medskip

\noindent
\textbf{Keywords:} Hepatitis C Virus, Deep Learning, Neural Network, Transformer Model
\end{abstract}

\newpage

\section{Introduction}
\label{sec:intro}

The virus responsible for hepatitis C (HCV) has been classified into 8 genotypes and more than 57 subtypes \cite{1, 2}. Treatment is generally carried out using direct-acting antivirals (DAAs) \cite{3}, and the choice of therapeutic regimen and treatment duration depends directly on the viral genotype, with emphasis on HCV subtypes 1a and 1b \cite{4}. Therefore, accurate viral genotyping directly influences the choice of treatment to be administered.

Currently, the amplification of genomic sequences by PCR, followed by sequencing and computational analysis, comprises the main methods for HCV genotyping \cite{5}, which include sequence alignment and phylogenetic analysis. In sequence alignment, tools such as the Basic Local Alignment Search Tool (BLAST) \cite{blast} allow comparison against reference sequences present in a database. In the construction of phylogenetic trees, an analysis of the level of similarity and evolutionary positioning of sequences is performed, enabling the identification of their viral genotype.

With the advancement of high-throughput sequencing (HTS) techniques \cite{novager}, which offer high speed and reduced costs in genetic sequencing, machine learning models represent a promising alternative for viral classification. Large language models (LLMs) based on Transformers, an architecture introduced by Google in the article \textit{Attention Is All You Need} \cite{transformers}, are particularly useful in natural language processing (NLP) tasks. The article describes the attention mechanism, which allows the model to assign greater importance to certain elements in a sequence, effectively "paying more attention". In traditional architectures, such as Recurrent Neural Networks (RNNs) and Long Short-Term Memory (LSTM), sentence processing is performed word by word, in a serial manner, whereas Transformers process all words in parallel. This allows these models to assign contextual weights to each element without relying on sequential processing, making training significantly faster and more efficient. This mechanism is called self-attention.

Two of the main representatives of Transformers are Generative Pre-trained Transformer (GPT) and Bidirectional Encoder Representations from Transformers (BERT) \cite{bert}. GPT is an autoregressive model, meaning it generates text by predicting the next word in a sequence based on the preceding context.
BERT, on the other hand, operates bidirectionally, considering both preceding and following context when predicting elements in a sequence.
More precisely, language models use tokens, which are basic units of text -- words, subwords, characters, or punctuation marks.
In general, GPT generates the next word (or the next token) based on what has been previously stated, whereas BERT predicts a word (or a token) within a sequence based on both prior and subsequent context. While GPT is more commonly used for translation and text generation, with ChatGPT \cite{chatgpt} as a prominent example, BERT models are more suitable for tasks such as sentiment analysis and language inference \cite{sentiment}.

DNABERT \cite{dnabert} is a recent example of how a Transformer-based LLM can be applied to the prediction of genetic sequences by interpreting them as sentences in natural language. Each sequence used during training undergoes a tokenization process, which creates a "dictionary" and assigns a corresponding numerical value to each token. In addition to this vocabulary, five special tokens are included to assist in the classification and training process using Masked Language Modeling (MLM). This technique masks specific tokens in a genetic sequence, and the model learns by making predictions about these tokens, which are then statistically evaluated.

In general terms, this study addresses the problem of alignment-free nucleotide sequence classification \cite{zielezinski2017}.
The rapid increase in nucleotide sequence data generated by high-throughput sequencing technologies demands efficient computational tools for sequence comparison.
Alignment-based methods, such as BLAST, are increasingly burdened by the scale of contemporary datasets due to their high computational demands for classification.

The use of machine learning methods to address the problem of alignment-free nucleotide sequence classification has shown considerable promise, especially for viral sequences \cite{wade2024}.
With the advent of Transformer neural networks, their application to nucleotide sequence classification problems is a very recent trend, and our study proposes a proof of concept \cite{servedio2014} for this type of neural network architecture.

The objective of this study is to explore the applicability of a Transformer-based neural network for the classification of genetic sequences, evaluating its performance in predicting HCV genotypes and subtypes after fine-tuning.

\section{Materials and methods}
\label{sec:metodo}

\subsection{Genetic data}
\label{sec:dadosgen}

The DNA sequences used in this study were obtained from the Los Alamos HCV Sequence Database \cite{hcv}, a public database that provides annotated HCV sequences.
The HCV dataset contains information on genotype, subtype, start and end coordinates relative to the HCV-H77 reference strain, country of sampling, city, date, and tissue associated with the sequences.

The inclusion criteria adopted in the search involved genotypes 1 to 6 (including recombinant variants) and all confirmed subtypes, considering only full-genome sequences. Genotypes 7 and 8 were excluded from the selection due to having only two sequences in the database, which could interfere with the model's fine-tuning stage.

In total, 2,881 HCV sequences were retrieved and used in this study, distributed across genotypes and subtypes shown in Table \ref{table:1}.

\begin{table}[!ht]
\caption{Distribution of HCV sequences across genotypes and subtypes.}
\centering
\begin{tabular}{lcccccccccccc}
\toprule
\textbf{Genotype} & \textbf{Total Count} & \textbf{a} & \textbf{b} & \textbf{c} & \textbf{d} & \textbf{f} & \textbf{g} & \textbf{h} & \textbf{i} & \textbf{k} & \textbf{u} & \textbf{v} \\
\midrule
1                 & 2,299                      & 1,666       & 629       & 4         & --         & --         & --         & --         & --         & --         & --         & --         \\
2                 & 301                       & 55         & 227       & 11        & --         & --         & --         & --         & 4         & 4         & --         & --         \\
3                 & 197                       & 190        & 5         & --         & --         & --         & --         & --         & --         & 2         & --         & --         \\
4                 & 29                        & 22         & --         & --         & --         & --         & --         & --         & --         & --         & --         & 7         \\
5                 & 10                        & 10         & --         & --         & --         & --         & --         & --         & --         & --         & --         & --         \\
6                 & 45                        & 23         & 2         & --         & 1         & 3         & 2         & 1         & --         & 5         & 4         & 4         \\
\bottomrule
\end{tabular}
\label{table:1}
\end{table}

\pagebreak

The generated files were processed using formatting scripts in order to assign a numerical label to each genotype or subtype of interest and to fragment the sequences. The fragment lengths varied across different datasets but remained close to 1,000.
This choice was based on a viral classification experiment conducted by the model's authors, which used SARS-CoV-2 sequences of length 1,000 to predict their variants.
Given that the HCV classification task is similar, we sought to adopt a comparable methodology.

For sequence fragmentation, three different scripts were developed, each applied exclusively to a specific dataset:

\begin{itemize}
    \item \texttt{seq\_break.py} -- Generates fragments of length 1,000. If fewer than 1,000 nucleotides remain at the end of the sequence, they are included as a smaller fragment.
    \item \texttt{new\_seq\_break.py} -- Generates fragments of length 999. Shorter trailing segments are discarded.
    \item \texttt{homo\_seq\_break.py} -- Divides sequences into 10 fragments of approximately equal length.
\end{itemize}

For each experiment, the selected sequences were distributed into three files: \textit{train} (training), \textit{dev} (validation), and \textit{test} (testing). Approximately 80\% of the data was allocated for training, with 10\% for validation and another 10\% for testing.

In the files used for fine-tuning with the suffix "...\_n", the distribution of genotypes or subtypes of interest within each file also followed an 80\%, 10\%, and 10\% proportion. This means that, for an example label with 100 samples total, 80 samples would be allocated to the training set, 10 to validation, and 10 to testing. In the remaining training setups, this proportion was applied to the total number of sequences, but not necessarily to each individual label.

\subsection{Model implementation}
\label{sec:modelo}

The machine learning model used was DNABERT-2 \cite{dnabert2}, a pre-trained Transformer-based model for classification and prediction of genetic sequences. Unlike DNABERT \cite{dnabert}, a previous model which uses fixed lengths per token created during training, DNABERT-2 builds a vocabulary of tokens with variable lengths based on frequent nucleotide segments. Combined with the MLM technique, this feature allows the prediction of both the length and the content of the masked token during training, increasing computational efficiency.

The experiments were carried out on a machine equipped with an NVIDIA GV100 GPU, with CUDA-accelerated processing.

Two modifications were made to the original code in order to facilitate fine-tuning and the application of the model after this stage:

\begin{itemize}
    \item Training script (\texttt{train.py}) -- adjustment of the saving paths for the model and tokenizer.
    \item Initialization script (\texttt{hcv.sh}) -- modification of directories and restructuring of the training steps into a single loop, enabling the use of a custom dataset.
\end{itemize}

Additionally, as a complement to the statistical performance evaluation metrics automatically computed, scripts were developed for use after fine-tuning based on the Data2Vec framework \cite{data2vec}. The scripts load the saved model and tokenizer and, based on them, classify input data according to genotype or subtype. This can be done with individual genetic sequences or multiple sequences in the form of a CSV dataset, provided that it follows the same format as the datasets used during training.

All scripts used for processing genetic data, training, and applying the model after fine-tuning, as well as additional results and comparisons between models, are available in the repository DNABERT-2\_HCV \cite{git}\footnote{\url{https://github.com/ariellaaro/DNABERT-2_HCV}}. The material also includes detailed instructions for implementing and applying the model to any dataset.

\subsection{Classification experiments}
\label{sec:typespred}

We performed two types of experiments: (i) genotype prediction and (ii) subtype prediction.

In the first case, sequences belonging to genotypes 1 to 6 were used, labeled accordingly with numerical values. Datasets were given distinct names and generated with different fragmentation methods, as shown in Table~\ref{table:2}.

\begin{table}[!ht]
    \centering
    \caption{Fine-tuning datasets used for genotype classification.}
    \begin{tabular}{ccc}
    \toprule
        \textbf{Name} & \textbf{Fragmentation} & \textbf{Genotypes} \\
        \midrule
        1train   & seq\_break      & 1 to 6 \\
        1train\_n & homo\_seq\_break & 1 to 6 \\
        2train   & new\_seq\_break  & 1 to 6 \\
        3train   & homo\_seq\_break & 1 to 6 \\
    \bottomrule
    \end{tabular}
\label{table:2}
\end{table}

For the second type of prediction, sequences belonging to the main subtypes---1a, 1b, 2a, 2b, and 3a---were used. The datasets were also given distinct names, but sequences were fragmented using one single strategy, making the comparison of results more consistent. Some of the datasets had specific characteristics regarding sequence distribution and labeling, as detailed in Table~\ref{table:3}. All subtypes were assigned numerical labels that remained consistent across all sequence datasets.

\begin{table}[!ht]
    \centering
    \caption{Fine-tuning datasets used for subtype classification.}
    \begin{tabular}{cccp{6cm}}
        \toprule
        \textbf{Name} & \textbf{Fragmentation} & \textbf{Subtypes} & \textbf{Notes} \\
        \midrule
        4train   & homo\_seq\_break      & 1a, 1b, 2a, 2b, 3a & dev and test sets are identical \newline and represent 20\% of the data \\
        5train   & homo\_seq\_break      & 1a, 1b, 2a, 2b, 3a &  \\
        5train\_n & homo\_seq\_break & 1a, 1b, 2a, 2b, 3a &  \\
        6train\_n & homo\_seq\_break & 1a, 1b, 2a, 2b, 3a & includes genotypes 4 to 6 \newline with a single label \\
        7train\_n & homo\_seq\_break & 1a, 1b, 2a, 2b, 3a & includes genotypes 4 to 6 \newline with a distinct label for each \\
        \bottomrule
    \end{tabular}
\label{table:3}
\end{table}

\section{Results}
\label{sec:result}

\subsection{Genotype classification}
\label{sec:genot}

Four fine-tuning runs were performed for HCV genotype classification. For each run, DNABERT-2 generates a report containing statistical measures computed at the final stage of training, referring to the evaluation dataset. In this study, loss, accuracy, and precision were considered for ranking and comparing the performance of the different training runs. The obtained results are shown in Table~\ref{table:4}.

\begin{table}[!ht] \centering \caption{Model performance comparison of fine-tuning runs for genotype classification.} \begin{tabular}{ccccccc} \toprule \textbf{Name} & \textbf{Loss} & \textbf{Accuracy} & \textbf{Precision} & \textbf{Recall} & \textbf{F1} & \textbf{MCC} \\ \midrule 1train & 0.077 & 98.471\% & 48.118\% & 55.324\% & 50.920\% & 0.909 \\ 1train\_n & 0.001 & \textbf{100\%} & \textbf{100\%} & 100\% & 100\% & 1.000 \\ 2train & \textbf{0.000} & \textbf{100\%} & \textbf{100\%} & 100\% & 100\% & 1.000 \\ 3train & 0.002 & 99.065\% & 98.485\% & 98.333\% & 98.329\% & 0.998 \\ \bottomrule \end{tabular} 
\label{table:4}
\end{table}

The script for prediction of individual sequences was applied using data from the test sets after fine-tuning the \texttt{1train} and \texttt{1train\_n} models. Based on the manual evaluation of the predicted labels, it was possible to identify a high error rate exclusively for the classification of genotype 5 sequences in the \texttt{1train} model, a result that was not observed in \texttt{1train\_n} or for other genotypes.

\subsection{Subtype classification}
\label{sec:subt}

Five fine-tuning runs were performed for HCV subtype classification. As with genotype classification, the model generates an evaluation report, with the results presented in Table~\ref{table:5}.

\begin{table}[h] \centering \caption{Model performance comparison of fine-tuning runs for subtype classification.} \begin{tabular}{ccccccc} \toprule \textbf{Name} & \textbf{Loss} & \textbf{Accuracy} & \textbf{Precision} & \textbf{Recall} & \textbf{F1} & \textbf{MCC} \\ \midrule 4train & 0.016 & 99.785\% & 99.677\% & 99.605\% & 99.641\% & 0.996 \\ 5train & 0.124 & 98.944\% & 99.048\% & 99.528\% & 99.273\% & 0.985 \\ 5train\_n & 0.010 & \textbf{99.891\%} & \textbf{99.871\%} & 99.905\% & 99.888\% & 0.998 \\ 6train\_n & \textbf{0.009} & 99.860\% & 99.436\% & 99.911\% & 99.669\% & 0.998 \\ 7train\_n & 0.011 & 99.860\% & 98.586\% & 99.690\% & 99.111\% & 0.998 \\ \bottomrule \end{tabular}
\label{table:5}
\end{table}

\subsection{Examples of model output}
\label{sec:aplic}

After fine-tuning, individual sequences and test datasets were provided to the model in order to predict their corresponding labels (genotypes or subtypes of interest).

\begin{figure}
    \centering
    \includegraphics[width=1\linewidth]{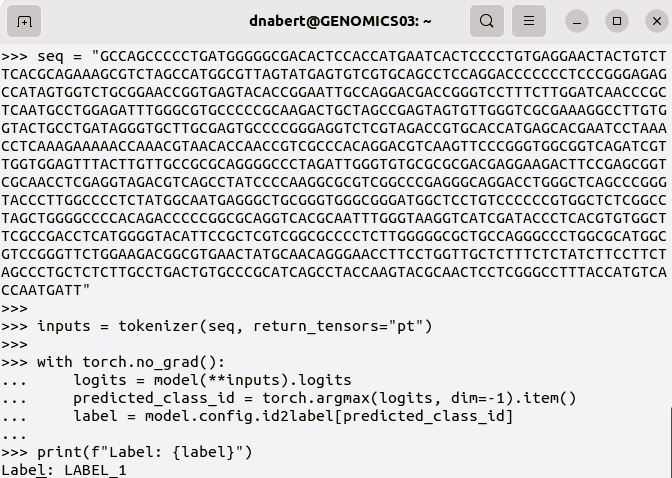}
    \caption{Individual HCV sequence prediction. Using the Python scripts available in the repository, users can provide a complete or partial HCV sequence as input, after which the model returns the corresponding label for the predicted classification.}
    \label{image1}
\end{figure}

Figure~\ref{image1} illustrates the execution environment of the developed scripts, showing an HCV sequence of genotype 1 and the predicted label (\texttt{LABEL\_1}), which corresponds to this genotype.

Figure 2 shows a portion of the file generated after providing a dataset with multiple test sequences, containing the predicted labels for each one. Correspondence verification between predicted labels and the true classification of the sequences was performed manually.

\begin{figure}
    \centering
    \includegraphics[width=1\linewidth]{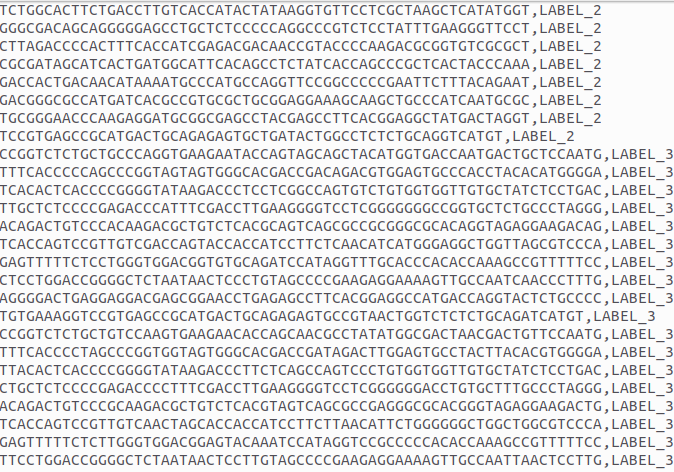}
    \caption{Multiple HCV sequence prediction. Using the Python scripts available in the repository, users can provide a file containing two or more HCV sequences as input, following the same format used for fine-tuning. The model returns the input sequences along with their corresponding labels for the predicted classifications.}
    \label{image2}
\end{figure}

\section{Discussion}
\label{sec:discuss}

\subsection{Genotype classification}
\label{sec:genot1}

In the genotype classification task, the model with the greatest performance discrepancy was \texttt{1train}. The low precision combined with high accuracy may be related to an inadequate sequence segmentation method, as well as an imbalanced distribution of labels within the datasets.

In this and other trainings that do not include the "...\_n" suffix, sequences were allocated arbitrarily among \textit{train}, \textit{dev}, and \textit{test} without a defined criterion regarding the proportion of each label per dataset. This may have caused class imbalance, with certain genotypes being heavily represented in one set and underrepresented in others. For example, if only one genotype 5 sequence is present in the training set and the remaining ones are allocated to the test set, the model may systematically fail to predict this genotype, reducing its precision. In addition, the combination of fragments of length 1,000 with considerably smaller ones may have interfered with the tokenization process, leading to an uneven allocation of tokens, making it more difficult to extract relevant and consistent patterns.

In subsequent models, \texttt{1train\_n} and \texttt{3train}, in which fragmentation and distribution were improved, an increase in both accuracy and precision was observed. Furthermore, after manual evaluation of the \texttt{1train\_n} model, sequences from the genotype with the highest error rate in \texttt{1train} were correctly classified. Thus, the improved performance suggests that correcting these factors enhanced the generalization capability of these models.

Although the \texttt{2train} model showed high performance, it was not considered suitable. This is due to the sequence splitting method used, which disregards nucleotides located at the end of the sequences. Since these regions may contain essential information for genotype identification, their exclusion compromises the integrity of the training data. Therefore, its evaluation metrics, despite being high, may be biased due to the loss of relevant information, limiting the model’s reliability in classifying new data.

Thus, for HCV genotype classification, the best-performing model was \texttt{1train\_n}, with sequences from genotypes 1 to 6 fragmented into segments corresponding to 1/10 of their original length and  evenly distributed across datasets.

\subsection{Subtype classification}
\label{sec:subt1}

For the subtype classification task, the \texttt{5train\_n} model achieved the best accuracy and precision results, due to its restricted classification scope (considering only subtypes, excluding additional genotypes) and the quality of the datasets, which followed the standard percentage distribution (80\%, 10\% and 10\%) and maintained a balanced proportion of labels in each file.

The \texttt{5train} model, although presenting slightly lower metrics than \texttt{4train}, can be considered superior. The only difference between the datasets used in both cases was the distribution of sequences between the \textit{dev} and \textit{test} files, which do not influence training but rather the evaluation of the models. In \texttt{4train}, these sets were identical and represented 20\% of all available data, whereas in \texttt{5train} they followed the standard 10\% split for each. The duplication of sequences across the two sets resulted in data leakage, potentially inflating the evaluation metrics of the \texttt{4train} model. Therefore, despite its higher accuracy and precision, \texttt{5train} presents greater reliability.

In addition to the most frequent subtypes, the \texttt{6train\_n} and \texttt{7train\_n} models also included genotypes 4 to 6. While in \texttt{6train\_n} these genotypes were grouped under a single label without distinction, in \texttt{7train\_n} they were assigned three distinct labels, which resulted in a slight decrease in precision. This suggests that model performance tends to decrease as data becomes more specific, although accuracy remained high.

Thus, for HCV subtype classification, the best-performing model was \texttt{5train\_n}, which included sequences from the most frequent subtypes (1a, 1b, 2a, 2b, 3a) evenly distributed across datasets, without including additional genotypes or subtypes.

\subsection{Limitations and future perspectives}
\label{sec:lim}

Although results demonstrated the effectiveness of DNABERT-2 in classifying HCV genotypes and subtypes, a few limitations and opportunities for improvement were identified.

The main limitation of the performed fine-tuning runs was the imbalance in the number of sequences per genotype.
It is common in classification tasks to encounter imbalanced datasets, that is, some classes (minority classes) contain substantially fewer samples than others \cite{chen2024}.
In our case, there was a predominance of genotype 1 and low representation of genotypes 4, 5, and 6. Although this imbalance reflects their global occurrence, it may compromise the model’s ability to generalize correctly to underrepresented classes. To mitigate this effect, strategies such as manually adjusting class weights or oversampling less frequent genotypes could be explored. Algorithms such as SMOTE \cite{smote} could potentially be explored to address this issue, although their application would require an appropriate sequence representation. In addition, expanding the dataset with sequences from other sources may contribute to more robust training.

Another aspect to be further explored is the sequence fragmentation strategy. Dividing sequences into overlapping blocks without a fixed ordering, for example, could improve the capture of relevant patterns and better distribute the available data, increasing the number of samples. Testing different segmentation approaches may lead to optimization of tokenization and, consequently, improve predictions.

To facilitate application after fine-tuning, the development of scripts that improve the model's usability could make it more practical in clinical and laboratory settings. One example would be the inclusion of a confidence metric for each prediction, allowing users to assess the robustness of classifications individually. Another useful feature would be the implementation of a label for sequences that do not correspond to any of the genotypes or subtypes present in training, facilitating the identification of novel data.

Finally, it is important to explore the performance of models other than DNABERT-2, since the use of Transformer-based architectures for genetic sequences is still recent and has advanced rapidly in recent years. Comparing different approaches may contribute to improving dataset construction strategies, as well as to a better understanding of their impact on training and model performance.

\section{Conclusion}
\label{sec:conclusao}

This study explored the application of Transformer-based models for nucleotide sequence classification, evaluating the performance of DNABERT-2 in predicting HCV genotypes and subtypes. Analyses considered accuracy and precision metrics after fine-tuning performed on different datasets.

The first experiment, conducted with inadequately fragmented sequences and uneven data distribution, resulted in a precision of 48.12\%. However, in subsequent training runs, with balanced datasets, precision increased to a range between 98.5\% and 100\%. This reinforces the importance of proper sequence preprocessing, a factor that directly influences model performance.

In addition, class imbalance may lead to systematic failures in predicting less frequent genotypes and subtypes. Frequently, genotype 5 sequences were classified incorrectly when the model was trained on an imbalanced dataset. This indicates the need for greater attention to sample distribution, as well as the exploration of methods to mitigate this bias, such as oversampling techniques or dataset expansion.

Thus, the results demonstrate that Transformer-based neural networks are a promising approach for viral sequence classification. They achieve high precision when trained with properly fragmented and well-distributed sequences and may, in the future, become an effective tool in clinical and laboratory settings.

\section*{Acknowledgments}
\noindent
The research of AA was supported by Funda\c{c}\~ao de Amparo \`a Pesquisa do Estado de S\~ao Paulo (FAPESP) grants 2020/08943-5 and 2023/14749-5; FA, MRSB and LMRJ were supported by Funda\c{c}\~ao de Amparo \`a Pesquisa do Estado de S\~ao Paulo (FAPESP) grants 2020/08943-5.
This article is based on AA’s undergraduate thesis, developed at the Escola Paulista de Medicina, Universidade Federal de São Paulo, under the supervision of FA.

\bigskip

\paragraph{Conflicts of interest.} None declared.


\bibliographystyle{abbrv}
\bibliography{refs}

\end{document}